\documentclass[journal,transmag]{IEEEtran}
\usepackage[utf8]{inputenc}

\usepackage[pdfencoding=auto, psdextra]{hyperref}
\usepackage[fleqn]{amsmath}
\usepackage{amsfonts}
\usepackage{amssymb}
\usepackage{eucal}
\usepackage[dvipsnames]{xcolor}
\usepackage{cite}
\usepackage{algorithm2e}
\usepackage{graphicx}
\usepackage[percent]{overpic}
\usepackage{bm}
\usepackage{ulem}

\newcommand{\Oinf}{\tilde{\mathcal{O}}_\infty}

\newtheorem{theorem}{Theorem}[section]
\newtheorem{remark}[theorem]{Remark}

\newtheorem{problem}[theorem]{Problem}
\newtheorem{definition}[theorem]{Definition}

\newtheorem{proposition}[theorem]{Proposition}

\SetKwRepeat{Do}{do}{while}

\title{Elimination of Redundant Polynomial Constraints and Its Use in Constrained Control}

\author{\IEEEauthorblockN{Andres Cotorruelo\IEEEauthorrefmark{1},
Ilya Kolmanovsky\IEEEauthorrefmark{2}, Daniel R. Ram\'irez\IEEEauthorrefmark{3}, Daniel Limon\IEEEauthorrefmark{3}, and
Emanuele Garone\IEEEauthorrefmark{1}}

\IEEEauthorblockA{\IEEEauthorrefmark{1}Service d'Automatique et d'Analyse des Syst\`emes, Universit\'e Libre de Bruxelles, Brussels, Belgium}
\IEEEauthorblockA{\IEEEauthorrefmark{2}Department of Aerospace Engineering, University of Michigan, MI, USA}
\IEEEauthorblockA{\IEEEauthorrefmark{3}Departamento de Sistemas y Autom\'atica, Universidad de Sevilla, Spain}
\thanks{This research has been funded partly by Ministerio  de  Econom\'ia  y  Competitividad  of Spain under project DPI2016-76493-C3-1-R co-financed by European FEDER Funds. The second author would like to acknowledge the support of National Science Foundation grant 1931738.}}

\begin{document}

\maketitle
\begin{abstract}
The reduction of constraints to obtain minimal representations of sets is a very common problem in many engineering applications. While well-established methodologies exist for the case of linear  constraints, the problem of how to detect redundant non-linear constraints is an open problem.
In this paper we present a novel methodology based on Sum of Squares for the elimination of redundant polynomial constraints. The paper also presents some relevant applications of the presented method to constrained control problems. In particular, we show how the proposed method can be used in the Model Predictive Control and in the Reference Governor frameworks to reduce the computational burden of the online algorithms. Furthermore, this method can also be used to eliminate the terminal constraints in MPC in a simple way that is independent from the cost function. 
\end{abstract}
\paragraph*{Notation}
The set of all polynomials with coefficients in $\mathbb{R}$ and variables $x_1,\ldots,x_n$ is denoted by $\mathbb{R}[x_1,\ldots,x_n]$. The set of all Sum of Squares (SOS) polynomials in variables $x_1,\ldots,x_n$ is denoted by $\Sigma[x_1,\ldots,x_n]$. The set of all nonnegative integers is denoted by $\mathbb{Z}_{\geq 0}$. Given two vectors $u$ and $v$, $(u,v)$ denotes $[u^T\ v^T]^T$. A set of elements $\{p_1,p_2,\ldots,p_n\}$ is denoted by $\{p_i\}_{i=1}^n$
\section{Introduction and Problem Statement}\label{sec:ps}
In many applications \cite{hillier2012introduction, gilbert1991linear,alessio2009survey,garone2015explicit}, engineers make use of sets described as intersections of inequalities  of the form
\begin{equation}\label{firstset}
\Omega = \{z \in \mathbb{R}^n : g_i(z) \geq 0,  i=1,...,n_c \},
\end{equation}
 where $n_c$ is the number of inequalities (\textit{i.e.} constraints). 

\begin{figure}[t]
    \centering
    \begin{overpic}[trim={.5cm .5cm .5cm .5cm},clip,width=.75\linewidth]{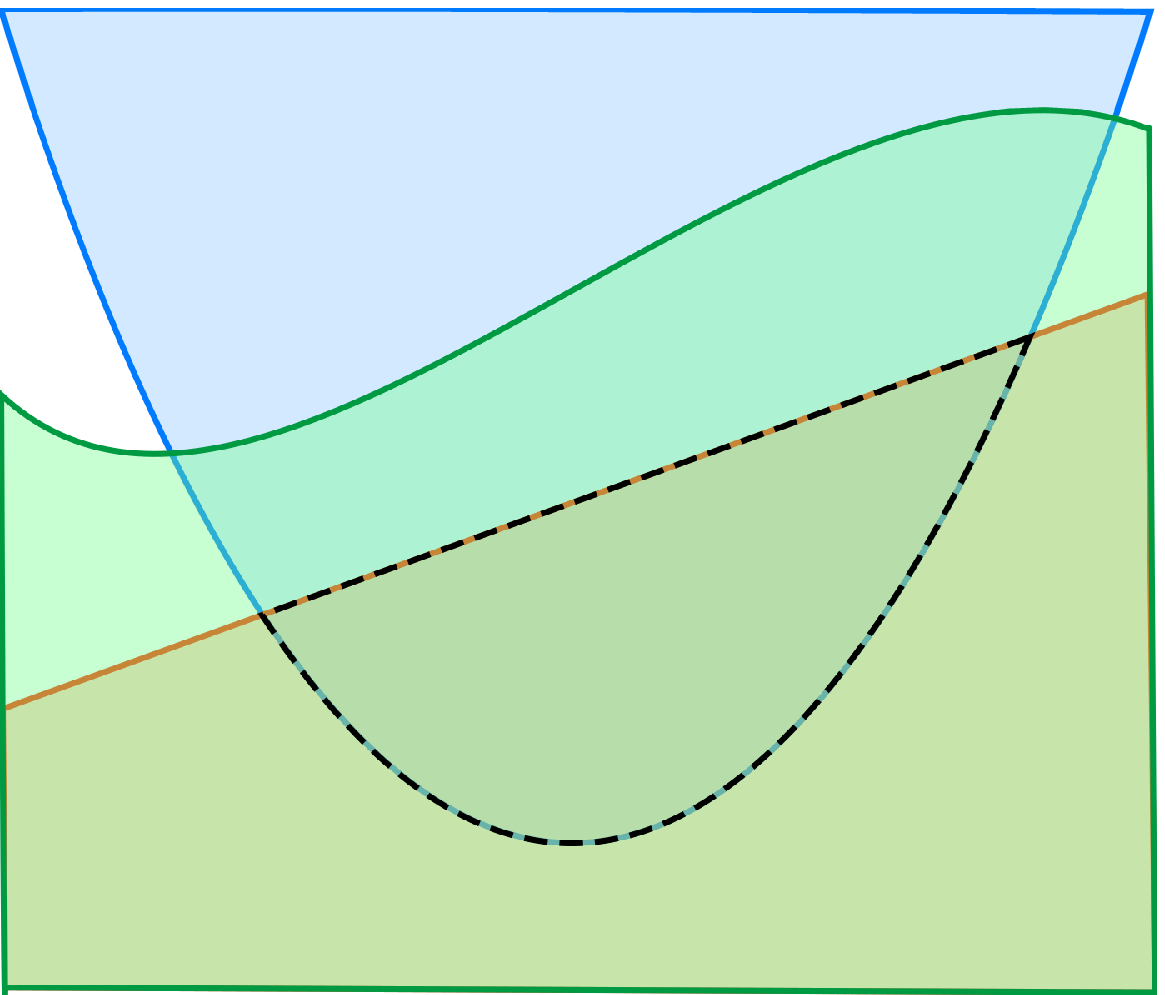}
    \put (20,5) {$g_2(z) \geq 0$}
    \put (65,60) {$g_3(z) \geq 0$}
    \put (15,65) {$g_1(z) \geq 0$}
    \put (50,25) {\Large$\Omega$}
    \end{overpic}
    \caption{The set $\Omega$ is defined as the intersection of three regions: $\{z:g_1(z)\geq 0\}$, $\{z:g_2(z)\geq 0\}$, and $\{z:g_3(z)\geq 0\}$. As it can be seen, $g_3(z)$ is redundant with respect to $g_1(z)$ and $g_2(z)$.}
    \label{fig:visual}
\end{figure}

Typically, the computational burden of algorithms that make use of these sets is a function of the number of constraints $n_c$, \cite{krupa2020implementation}. As a consequence, in many contexts, it is highly desirable to use \textit{minimal representations} of sets, i.e. sets where none of the constraints is redundant. We recall that a constraint $c(z) \geq 0$ is redundant \textit{w.r.t.} a set $\Omega$ if 
$
\Omega \cap \{z\in \mathbb{R}^n : c(z) \geq 0\} = \Omega.
$
\figurename~\ref{fig:visual} gives an example of a set defined using three constraints of which one is redundant.

In this paper we focus on the problem of determining if a  constraint is redundant, and thus can be eliminated.
Applications where the ability to perform redundant constraint elimination is very useful include the minimal representation of Maximal Output Admissible Sets (MOAS) \cite{tan1992maximal},  synthesis of Petri net supervisors \cite{huang2015synthesis}, analysis of power systems \cite{hua2013eliminating}, and even Mendelian genetics \cite{xiao2007fast}. It must be also noted that redundant constraints elimination is at the basis of some inner and outer set approximations techniques, see \textit{e.g.} \cite{gilbert1999fast}.
More formally, in this paper we focus on the following problem:

\begin{problem}{(Redundancy certificate)}\label{problemStatement}
Determine a certificate that guarantees that a constraint $c(z) \geq 0$ is redundant \textit{w.r.t.} to a set $\Omega$.
\end{problem}

A well-known method to obtain a redundancy certificate is to use  the fact that a constraint $c(z)>0$ is redundant \textit{w.r.t.} $\Omega$ if and only if $c(z) \geq 0, ~\forall z \in \Omega$. Accordingly it is enough to compute
\begin{equation}\label{optim}
    \rho^* = \min_{z\in \Omega} c(z)
\end{equation}
and to note that $\rho^* \geq 0$ if and only if the constraint is redundant. The main difficulty with this formulation is that it provides a redundancy certificate only in the case one can compute the actual optimal solution of \eqref{optim}. If instead one can compute only a suboptimal solution $\rho \leq \rho^*,$ the only thing that can be concluded is non-redundancy in the case $\rho<0,$ otherwise nothing can be said. Accordingly, the above approach can be actually used to eliminate redundant constraints only when it is reasonable  to solve exactly problem \eqref{optim}, \textit{e.g.} when \eqref{optim} is convex. Note that for \eqref{optim} to be convex the constraint to be checked must be concave and the set $\Omega$ needs to be convex. For this reason this methodology is used almost exclusively to test the redundancy of linear constraints \textit{w.r.t.} polyhedral sets.

Based on this idea, several algorithms have been proposed in the literature to determine redundancy for linear constraints. Approaches range from Linear Programming \cite{garone2017reference,caron1989degenerate}, to heuristic  \cite{paulraj2006heuristic}, and to deterministic methods \cite{telgen1983identifying}. For an extensive survey on the subject, the reader is referred to \cite{paulraj2010comparative}. Other notable papers on redundant constraints elimination include  \cite{scholl2009computing} where the authors use Satisfiable Modulo Theories to perform redundant linear constraint elimination for non-convex polyhedra, and  \cite{zou2016redundant} where the authors use Path Condition arguments to detect redundancy in the framework of symbolic execution.

To the best of the authors' knowledge, the existing literature does not provide any systematic methods for the determination of redundant nonlinear constraints. In this paper we propose a novel procedure that allows to ascertain whether a polynomial constraint is redundant with respect to a set defined by polynomial constraints. The proposed solution makes use of the Sum of Squares framework.

The elimination of redundant constraints is particularly desirable in constrained control, where constrained optimization problems need to be solved in real time \cite{rawlings2009model,garone2017reference}. For this reason in this paper we will also show some relevant applications of this methodology to constrained control. 

\section{Construction of the Certificate}\label{sec:methodology}
Consider Problem
\ref{problemStatement} where the set $\Omega$ is in the form 
\eqref{firstset}. We assume that the functions $c(z),~g_i(z)$ are polynomials, i.e.  $c(z),~g_i(z)\in\mathbb{R}[z], i=1,...,n_c$. 

To build a redundancy certificate we will make use of  the Krivine--Stengle Positivstellensatz (P-satz) \cite{stengle1974nullstellensatz, parrilo2000structured}, see Appendix \ref{sec:psatz}. The first step to work with the P-satz is to express the conditions of the certificate in terms of \textit{set emptiness}. In the case at hand, the problem can be expressed as
\begin{equation}\label{eq:set0}
 \left\{z:z\in\Omega,\, c(z)<0 \right\}=\emptyset.
\end{equation}

Since the Krivine--Stengle P-satz is defined for \textit{greater-than-or-equals-to}, \textit{equals-to}, and \textit{not-equals-to} operators, the set \eqref{eq:set0} must be reformulated as
\begin{equation}\label{eq:set1}
    \left\{z\!:g_1(z)\!\geq\!0,\ldots,g_{n_c}(z)\!\geq\! 0\right.,
    \left.c(z)\geq0,\,c(z) \!\neq \! 0 \right\}\!=\!\emptyset.
\end{equation}

Applying the Krivine--Stengle P-satz, \eqref{eq:set1} becomes equivalent to the existance of two polynomials $a,b$ such that \footnote{In the sequel, for the sake of notational clarity the arguments of functions will be left out when there is no risk of confusion.}
\begin{equation}
    a\!+\!b^2\!=\!0,~a\in\textnormal{Cone}\left(\left\{-c,g_1,\ldots,g_n \right\}\right),~b\in\textnormal{Monoid}(c).
\end{equation}

By performing standard algebraic manipulations, it results that if there exists Sum of Squares polynomials $s_i \in \Sigma[z], i=0,...,n_c$ such that
\begin{equation}\label{eq:cond}
    -s_0 c - \sum_{i=1}^{n_c} c s_i g_i + c^2=0,
\end{equation}
then \eqref{eq:set1} is true, \textit{i.e.} $c(z)\geq 0$ is redundant \textit{w.r.t.} $\Omega$.
Condition  \eqref{eq:cond}
can be further simplified by putting $c$ in evidence which leads to
\begin{equation}\label{eq:cond2}
    -s_0 - \sum_{i=1}^{n_c} s_i g_i + c=0.
\end{equation}
Since $s_0 \in \Sigma[x]$ this condition can be rewritten as
\begin{equation}\label{eq:cond3}
     c - \sum_{i=1}^{n_c} s_i g_i \in\Sigma[z].
\end{equation}

Equation \eqref{eq:cond3} implies that, if there exist $s_1,\ldots,s_n \in \Sigma [z]$ such that $c - \sum_{i=1}^n s_i g_i$ is a sum of squares polynomial, $c(z)\geq 0$ is redundant \textit{w.r.t.} $\Omega$. This condition can be checked using the following SDP feasibility test

\arraycolsep=1.4pt
\begin{equation}\label{eq:opt_alpha}
\begin{array}{lrll}
      \textrm{find } s_i,\, i=1,\ldots,n_c&&&\\
\text{s.t.}&&&\\
&c - \sum_{i=1}^{n_c} s_i g_i &\in\Sigma[z]\\
&s_i&\in\Sigma[z],~i=1,\ldots,n_c.
\end{array}
\end{equation}

If \eqref{eq:opt_alpha} is feasible, then $c(z)\geq 0$ is redundant \textit{w.r.t.} $\Omega$.
In many cases it is also of interest to  quantify ``how much" a constraint is redundant. A possible way to do so is to maximize on a slack variable $\rho$ as follows

\begin{equation}\label{eq:opt_rho}
\begin{array}{llrl}
\rho^*= &\max \rho &\\
        &\text{s.t}&\\
        &          &c  - \sum_{i=1}^{n_c} s_i g_i - \rho &\in\Sigma[z]\\
        &          &s_i                             &\in\Sigma[z],\,i=1,\ldots,n_c.
\end{array}
\end{equation}
In this formulation, if $\rho^*$ is positive the constraint is redundant. This slack variable approach can also be used to assess which constraints are ``almost redundant" and can be possibly eliminated using inner constraints approximations like the ones resulting of the pull-in transformation presented in  \cite{gilbert1999fast}.

Note that the optimization problem \eqref{eq:opt_rho} has $\binom{n+d_s}{n}n_c+1$ decision variables,  where $d_s$ is the chosen degree of the $s_i$, and $n$ is the number of variables of the constraints. The problems has $n_c$ LMI conditions, each of size $\binom{n+\bar{d}_s}{n} \times \binom{n+\bar{d}_s}{n}$, with $\bar{d}_s=\frac{d_s}{2}$, and one LMI condition of size $\binom{n+\bar{d}_g}{n} \times \binom{n+\bar{d}_g}{n}$, with $\bar{d}_g=\lceil \frac{d_s+d_g}{2} \rceil$, where $d_g$ is the maximum degree of the $g_i$. This problem can be efficiently solved by available Sum of Squares Programming optimization software \textit{e.g.} \cite{Lofberg2004}. 

\section{Applications to Constrained Control}
In this section, we illustrate some examples of possible uses of 
the proposed redundant constraint elimination procedure in constrained 
control.

\subsection{Constraint elimination in Model Predictive Control}
Model Predictive Control (MPC) is one of the most successful advanced control schemes both in industry and academia \cite{hrovat2012development}. The main idea behind MPCs is to compute a sequence of control actions $\textbf{u}$ at every time step by solving an optimization problem.

A typical MPC computes the control sequence over a control horizon $N_c$, and assumes that the system is controlled with a terminal control law $\kappa(x)$ for the rest of the prediction horizon $N_p \geq N_c$. Moreover, the last explicitly predicted state is usually constrained to belong to a  terminal set to  guarantee recursive feasibility and  stability \cite{maciejowski2002predictive}. 

Due to the need of solving an optimization problem at each time step, it is easy to see the advantage in being able to remove redundant constraints to reduce both the online computational burden and memory usage. 

Usually the problem solved by an MPC at each time step is defined as  an optimization problem  parameterized in  the initial state $x_0$ and where the decision variables are aggregated into the input sequence $\textbf{u}$
  \begin{IEEEeqnarray}{ll}
    \min_\textbf{u} J(\textbf{u},x_0) \IEEEyesnumber \IEEEyessubnumber \label{eq:mpc_cost}\\
    \textnormal{s.t.} \nonumber\\
    c_{u,i}\left(\hat{u}(j|x_0,\textbf{u},\kappa(\cdot))\right)\geq 0 \quad &i=1,\ldots,n_{c,u}\IEEEyessubnumber \label{eq:mpc_first}\\ &j=0,\ldots,N_p-1  \nonumber \\
    c_{x,i}\left(\hat{x}(j|x_0,\textbf{u},\kappa(\cdot))\right) \geq 0\quad &i=1,\ldots,n_{c,x} \IEEEyessubnumber \\  &j=0,\ldots,N_p-1 \nonumber \\
    \hat{x}(N_p|x_0,\textbf{u},\kappa(\cdot)) \in \Omega_t, \IEEEyessubnumber \label{eq:mpc_terminal}
\end{IEEEeqnarray}
where $c_{u,i}(u)\geq0$  and $c_{x,i}(x)\geq0$ are the control and state constraints of the system, $\Omega_t$ is the terminal set, and 
$\hat{u}(j|x,\textbf{u},\kappa(\cdot))$ and $\hat{x}(j|x,\textbf{u},\kappa(\cdot))$ are the predictions of the input and of the state, respectively, defined under the assumption that the input is 
$\textbf{u}(j)$ for the first $N_c$ steps and is generated by the control law $\kappa(\cdot)$ for the last $N_p-N_c$ steps. 

For the goal of eliminating redundant constraints, a convenient  way to rewrite \eqref{eq:mpc_cost}-\eqref{eq:mpc_terminal} is
  \begin{IEEEeqnarray}{ll}
    \min_{\textbf{u}} J(\textbf{u},x_0) \label{eq:mpc_cost2}\\
    \textnormal{s.t.} \nonumber\\
(x_0,\textbf{u}) \in \Omega
\end{IEEEeqnarray}
where
\begin{equation}\label{OmegaMPC}
\Omega = \left\{(x,\textbf{u})\left|
\begin{array}{l}
c_{u,i}\left(\hat{u}(j|x,\textbf{u},\kappa(\cdot))\right)\geq 0,\\
\hfill i=1,\ldots,n_{c,u},\\ \hfill j=0,\ldots,N_p-1  \\ c_{x,i}\left(\hat{x}(j|x,\textbf{u},\kappa(\cdot))\right) \geq 0\\
\hfill i=1,\ldots,n_{c,x} \\ \hfill j=0,\ldots,N_p-1 ,\\ 
\hat{x}(N_p|x,\textbf{u},\kappa(\cdot)) \in \Omega_t,
\end{array}\right.
\right\}
\end{equation}

The main interest of  this formulation is that the dependency on the initial state is considered as a variable rather than as a parameter. This allows to  use the proposed approach to  eliminate redundant constraints in \eqref{OmegaMPC}, which in turn allows to eliminate redundant constraints in \eqref{eq:mpc_first} -- \eqref{eq:mpc_terminal}.

\subsubsection*{Application 1: Detection of Redundant Constraints}
Consider the following linear model of a ball-and-plate system obtained under the assumption that the plate inclination control is much faster than the dynamics of the ball:
\begin{equation}\label{eq:mpc_ap1}
    x(t+1)=\begin{bmatrix}
        1 & 0.5&    0&         0\\
         0 &   1       & 0     &    0\\
         0      &   0   & 1 &0.5\\
         0   &      0     &    0   & 1
    \end{bmatrix}x(t) + \begin{bmatrix}
        0.125   &      0\\
    0.5    &     0\\
         0  &  0.125\\
         0  &  0.5\\
    \end{bmatrix} u(t),
\end{equation}
where the state vector $x=[p_x~v_x~p_y~v_y]^T$ aggregates the horizontal and vertical position and velocity of the ball, and the control action $u=[u_x~u_y]$ is the inclination of the plate with respect to the horizontal and vertical axes. 
This system is subject to the following input and state constraints
\begin{equation}\label{eq:mpc_constraints}
    \begin{aligned}
    -x_1^4-x_3^4+10x_1^2-x_3^2+0.1&\geq 0,\\
    |x_2|\leq 2,\,|x_4|\leq 2,\,|u_1|\leq 2,\, |u_2|&\leq 2.
    \end{aligned}
\end{equation}

\begin{figure}
    \centering
    \includegraphics[width=\linewidth]{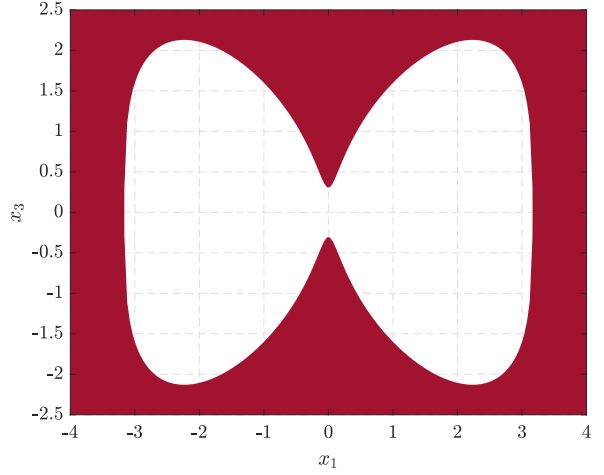}
    \caption{Depiction of the \textit{bow tie set}.}
    \label{fig:bowtie}
\end{figure}

The first constraint in \eqref{eq:mpc_constraints} forces the ball to stay within the \textit{bow tie set}, depicted in \figurename~\ref{fig:bowtie}. The remaining constraints are ball velocity constraints and input saturation constraints.
The system is controlled with an MPC in the form  \eqref{eq:mpc_cost} -- \eqref{eq:mpc_terminal}
 where $N_c=2$, $N_p=10$ and the terminal set is given by $\Omega_t=\{x:x^T P x \leq 1 \}$, with
 \begin{equation*}
     P=\begin{bmatrix}
     4.035  &  2.0616  &  0  &  0\\
    2.0616  &  4.1438  &  0  &  0\\
    0  &  0  &  4.035 &   2.0616\\
    0  &  0  &  2.0616  &  4.1438
     \end{bmatrix},
 \end{equation*}
 resulting in a total of 96 constraints.

By applying the constraint reduction algorithm presented in this paper it is possible to show that among these 96 constraints, 64 are redundant, reducing the total number of constraints of the MPC optimization problem by upwards of 67 \%. In Table \ref{tab:my_label} we report the average elapsed time over 1000 random MPC problems both when all constraints are considered, and when redundant constraints are eliminated. Note that although the computational time does not decrease linearly with the number of constraints, the required memory allocation does.

The constraint reduction  optimization problems were solved using MATLAB R2019b and YALMIP \cite{Lofberg2004}, running on an Intel Core i7-7500 at 2.7 GHz with 16 GB of RAM. The elapsed time for the redundancy test was 5 h, averaging 190 s to check each constraint.

\subsubsection*{Application 2: Elimination of the Terminal Constraint}
A special kind of a constraint whose elimination is very interesting is the terminal constraint. In \cite{limon2006stability} it was proven that it is possible to determine a region of the state space in which the terminal constraint is always satisfied by appropriately weighting the terminal cost. In \cite{hu2002toward}, the authors propose a non-standard terminal cost, such that stability is proven for a subset of the admissible states without the need for a terminal constraint. Methods based on relaxed dynamic programming inequalities \cite{grune2017nonlinear} can also be used to determine sufficiently long horizon to achieve closed-loop stability without terminal constraints; but their application is not straightforward.

In the framework presented in this paper it is natural to wonder \textit{when the terminal constraint is redundant with respect to the other constraints} and can thus be eliminated. It is possible to answer this question using the methodology presented in Section \ref{sec:methodology}. This is particularly useful when the system and the constraints are linear and the cost function is quadratic. In this case eliminating the quadratic terminal constraint would yield a Quadratic Programming problem.

In this example we will determine the smallest $N_p$ such that the terminal constraint is redundant for system the following system,
\arraycolsep 5pt
\begin{equation}\label{eq:mpc_system}
    x(t+1)=\begin{bmatrix}1 & 0.5\\
    0 & 1
    \end{bmatrix}x(t) + \begin{bmatrix}0.125\\
    0.5
    \end{bmatrix}u(t),
\end{equation}
subject to the following state and control constraints,
\begin{equation}\label{eq:mpc_const2}
 |x_1| \leq 5, \,\, |x_2| \leq 5, \,\, |u| \leq 1,
\end{equation}
and with the terminal constraint 
\begin{equation}\label{eq:mpc_example_terminal}
    x(N_p)^T\begin{bmatrix}
    4.0350  &  2.0616\\
    2.0616  &  4.1438
    \end{bmatrix}x(N_p)\leq 10.
\end{equation}
Assuming a control horizon $N_c=3$, we computed the smallest $N_p$ such that \eqref{eq:mpc_example_terminal} becomes redundant with respect to the other constraints, which is $N_p=7$. This means that whenever this system is controlled with an MPC with $N_p\geq 7$, the terminal constraint is automatically fulfilled by the other constraints and can be removed, thus transforming the problem from a QPQC to a QP. On top of this, it is also possible to prove that 36 out of the total 46 constraints are  redundant, reducing the total number of constraints by upwards of 78\%. Computational times are reported in Table \ref{tab:my_label} for two different QP solvers.

\begin{remark}
An alternative approach to remove the terminal set is to perform vertex enumeration on the set \eqref{OmegaMPC} excluding the terminal constraint, and checking if every vertex is inside the terminal set. However, the complexity of vertex enumeration algorithms grows exponentially in the dimension of the state and in the number of the constraints \cite{khachiyan2009generating} while the proposed approach scales polynomially.
\end{remark}

\begin{table*}[ht]
    \centering
    \begin{tabular}{|c|c|c|c|c|}
       \hline System  & Algorithm & Avg. time w\slash o constraint elimination & Avg. time w\slash~ constraint elimination & Reduction \\
       \hline \eqref{eq:mpc_ap1} -- \eqref{eq:mpc_constraints} & MATLAB's \texttt{fmincon} & 21 ms & 18.7 ms & 11 \% \\ \hline
       \eqref{eq:mpc_system} -- \eqref{eq:mpc_example_terminal} & MATLAB's \texttt{quadprog} & 1.6 ms & 1.4 ms & 14.5 \% \\ \hline
        \eqref{eq:mpc_system} -- \eqref{eq:mpc_example_terminal} & MOSEK's \texttt{quadprog} & 4.5 ms & 4 ms & 10.2 \% \\ \hline
    \end{tabular}
    \vspace{.5 cm}
    \caption{Comparison of the computational times of the MPC optimization problem}
    \label{tab:my_label}
\end{table*}

\subsection{Constraint elimination in Reference Governors}
The term Reference Governor (RG) \cite{kolmanovsky2014reference,garone2017reference} denotes a family of   control schemes based on the idea of decoupling the stabilization of the system from the satisfaction of constraints. In these schemes, the system is stabilized by means of a \textit{primary control law} $\kappa(x,v)$ designed so that the output of the system tracks an auxiliary reference $v$. This auxiliary reference is  manipulated by the RG so that at every time step it approximates as much as possible the reference defined by the user, typically denoted by $r,$ without violating the constraints. 

Typically, the way to check whether the constraints will be fulfilled for a given state $x$ and specified reference $v$ is by using the \textit{Maximal Output Admissible Set}:

\begin{definition}{(Maximal Output Admissible Set)}
Consider a precompensated system
\begin{equation}\label{eq:rg_system}
    x(t+1)=f(x(t),v(t)),
\end{equation}
subject to constraints
$ (x,v)\in\mathcal{D}. 
$
The MOAS is defined as
\begin{equation*}
    \mathcal{O}_\infty = \{(x,v):(\hat{x}(k|x,v),v)\in \mathcal{D},\,\forall k\in\mathbb{Z}_{\geq0}\},
\end{equation*}
where $\hat{x}(k|x,v)$ is the prediction of the state at time $k$ with initial state $x$ and constant applied reference $v$.
\end{definition}
In practice, it is common to use $\Tilde{\mathcal{O}}_\infty$, a slightly tightened version of $\mathcal{O}_\infty$ denoted as
\begin{equation*}
    \Tilde{\mathcal{O}}_\infty=\mathcal{O}_\infty \cap \mathcal{O}^\varepsilon,
\end{equation*}
with
$
    \mathcal{O}^\varepsilon=\{(x,v):(\overline{x}_v,v)\in(1-\varepsilon) \mathcal{D}\},
$ where $\overline{x}_v$ is the steady-state associated to the applied reference $v$ and $\varepsilon>0$ is a small constant. It is possible to prove \cite{gilbert1991linear} that under reasonable additional assumptions $\Tilde{\mathcal{O}}_\infty$ is \textit{finitely determined}, \textit{i.e.}, there exists a $k^\ast$ such that
\begin{equation*}
    \Oinf\!=\!\tilde{\mathcal{O}}_{k^\ast}\!=\!\{(x,v)\!:\!(\hat{x}(k|x,v),v)\in \mathcal{D}, \forall k\!=\!1,\ldots,k^\ast\}\cap\mathcal{O}^\varepsilon\!.
\end{equation*}
Under some further reasonable assumptions and using similar arguments to those in \cite{bemporad1998reference} it is possible to determine an upper bound of $k^\ast$, $\bar{k}\geq k^\ast$. However this upper bound is typically very conservative.

In \cite{gilbert1991linear} an iterative algorithm able to determine  $\Tilde{\mathcal{O}}_\infty$ (Algorithm \ref{alg:algorithm}) and $k^*$ was presented. The insight behind this algorithm is that it can be proven that $k^*$ is the smallest $k$ such that  $\Tilde{\mathcal{O}}_k^*=\Tilde{\mathcal{O}}_{k^*-1}.$ 

Note that in Algorithm \ref{alg:algorithm} the condition $\Tilde{\mathcal{O}}_k\neq\Tilde{\mathcal{O}}_{k-1}$ requires the capability of checking redundant constraints: if all newly added constraints are redundant with respect to $\Tilde{\mathcal{O}}_{k-1}$, then $\Tilde{\mathcal{O}}_k=\Tilde{\mathcal{O}}_{k-1}$ and the algorithm can stop.

\begin{algorithm}[ht]
$\Tilde{\mathcal{O}}_0 \leftarrow \mathcal{O}_0 \cap \mathcal{O}^\varepsilon$\\
$k\leftarrow0$\\
\Do{$\Tilde{\mathcal{O}}_k\neq\Tilde{\mathcal{O}}_{k-1}~\vee~k\leq \bar{k}$}{
$k \leftarrow k+1$\\
$\Tilde{\mathcal{O}}_k \leftarrow \Tilde{\mathcal{O}}_{k-1} \cap \{(x,v):c_i(\hat{x}(k|x,v),v)\geq0,i=1,\ldots,n_c\}$\\
}
$k^\ast\leftarrow k-1$
\caption{Computation of $\Tilde{\mathcal{O}}_\infty$}
\label{alg:algorithm}
\end{algorithm}

However, as mentioned in the introduction, available procedures to determine if a constraint is redundant made this approach viable only for linear systems subject to linear constraints. Indeed, in most papers so far, Reference Governors for nonlinear systems and/or nonlinear constraints have used either a very conservative $\bar{k}$ derived by Lyapunov arguments or empirically estimated horizons. Note that in Reference Governors the computational effort is directly proportional on the number of constraints \cite{garone2017reference,bemporad1998reference}. With the approach proposed in this paper, it is finally possible to check constraint redundancy for polynomial constraints and thus use Algorithm \ref{alg:algorithm}.

Note that using Algorithm \ref{alg:algorithm} and the certificate proposed in Section \ref{sec:methodology} to compute $\Tilde{\mathcal{O}}_\infty$ requires solving $(k^\ast+1)\cdot n_c$ LMI feasibility tests as in \eqref{eq:opt_alpha}. Accordingly every iteration $k$ will take longer than the previous one, since the number of inequalities describing $\tilde{\mathcal{O}}_k$ grows by up of $n_c$ every iteration. Furthermore, also the number of variables increases as we need to declare a new SOS multiplier $s_i$ for every new inequality.  Interestingly, it is possible to reduce the number of redundancy checks and consequently the computational time by using the following proposition.

\begin{proposition}\label{lem:lemma}
If a constraint $c_j(x,v)\geq0$ is redundant at iteration $k$ \textit{w.r.t.} $\Tilde{\mathcal{O}}_k,$ then it will be redundant for any iterations $k'>k$.
\end{proposition}
PROOF - Let $\mathcal{C}_i=\{(x,v):c_i(x,v)\geq0\}$, $i=1,\ldots,n_c$ and let us assume that constraint $j$ becomes redundant at iteration $k'$. By definition
\begin{equation}
    (\hat{x}(k|x,v),v)\in\bigcap_{i=1}^{n_c} \mathcal{C}_i,\, k=\{1,\ldots,k'-1\},
\end{equation}
and since the $j$-th constraint is redundant at iteration $k'$
\begin{equation*}
  (\hat{x}(k'|x,v),v)\in\mathcal{C}_j.
\end{equation*}
Then for any $(x,v)\in\Tilde{\mathcal{O}}_{k'-1}$, $(\hat{x}(k'|x,v),v)\in\mathcal{C}_j$. Finally, since $\tilde{\mathcal{O}}_{k'}\subseteq\tilde{\mathcal{O}}_{k'-1}$, $(x,v)\in \mathcal{O}_{k'}$ implies $(f(x,v),v)\in\mathcal{C}_j$, therefore, the $j$-th constraint is redundant at iteration number $k'+1$.
$\hfill \blacksquare$

Using Proposition \ref{lem:lemma} we can refine Algorithm \ref{alg:algorithm} into the more computationally efficient Algorithm \ref{alg:algorithm2} which only checks the redundancy of constraints that have not been redundant so far.

\begin{algorithm}[ht]
$\Tilde{\mathcal{O}}_0 \leftarrow \mathcal{O}_0 \cap \mathcal{O}^\varepsilon$\\
$k\leftarrow0$\\
$\rho_i\leftarrow 0,\,i=1,\ldots,n_c$\\
\Do{$\bigwedge_{i=1}^{n_c}\rho_i \neq 1~\vee~k\leq \bar{k}$}{
    $k \leftarrow k+1$\\
    $\Tilde{\mathcal{O}}_k \leftarrow \Tilde{\mathcal{O}}_{k-1}$\\
    \For{$i=1,\ldots,n_c,\,\rho_i\neq1$}{
        Check redundancy of $c_i(\hat{x}(k|x,v),v)$ \textit{w.r.t.} $\tilde{\mathcal{O}}_{k}$\\
        \eIf{$c_i(\hat{x}(k|x,v),v)$ is redundant}{
        $\rho_i\leftarrow 1$
        }{
        $\Tilde{\mathcal{O}}_k \leftarrow \Tilde{\mathcal{O}}_k \cap \{(x,v):c_i(\hat{x}(k|x,v),v)\geq0\}$\\
        }
    }
}
$k^\ast\leftarrow k-1$
\caption{Improved Computation of $\tilde{\mathcal{O}}_\infty$}
\label{alg:algorithm2}
\end{algorithm}
\subsubsection*{Application 1: Electromagnetically Actuated Mass-Spring Damper}

The presented methodology is used to compute $\Oinf$ for an electromagnetically actuated mass-spring damper system \cite{gilbert2002nonlinear}, depicted in \figurename~\ref{fig:example}. This system is modeled by the following equations:
\begin{figure}[t]
    \centering
    \begin{overpic}[width=0.7\linewidth]{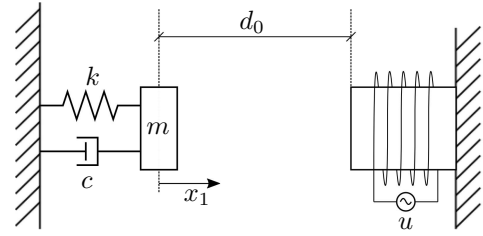}
    \put (15,35) {$k$}
    \put (14,13) {$c$}
    \put (28,24.5) {$m$}
    \put (34,10) { $x_1$}
    \put (46,47) { $d_0$}
    \put (80,4) { $u$}
    \end{overpic}
    \caption{Electromagnetically actuated mass-spring damper system.}
    \label{fig:example}
\end{figure}

\begin{equation}\label{eq:sys_example}
\begin{aligned}
\dot{x}_1&=x_2\\
\dot{x}_2&=-\frac{k}{m}x_1-\frac{c}{m}x_2+\frac{\alpha}{m}\frac{u}{(d_0-x_1)^\gamma},
\end{aligned}
\end{equation}
where $x_1$ and $x_2$ are the position and velocity of the armature, respectively, $k=38.94$ N m$^{-1}$, $m=1.54$ kg, $c=0.659$ N s m$^{-1}$, $\alpha=4.5\cdot10^{-5}$ C$^2$m$^{-3}$kg$^{-1}$, $d_0=0.0102$ m, and $\gamma=2$.
This system can be feedback linearized using the control law
\begin{equation}\label{eq:rg_u}
 u=\frac{1}{\alpha}(d_0-x_1)^\gamma (k v -c_d x_2),
\end{equation}
with $c_d=4$. The closed loop system becomes
\begin{equation}\label{eq:cl_system}
\begin{aligned}
\dot{x}_1&=x_2\\
\dot{x}_2&=-\frac{k}{m}x_1-\frac{c+c_d}{m}x_2+\frac{k}{m}v.
\end{aligned}
\end{equation}

System \eqref{eq:cl_system} is  subject to the following constraints:
\begin{equation*}
    x_1\leq 0.008,\quad k v - c_d x_2 \geq 0, \quad u\leq 0.3,
\end{equation*}
where $u$ is given by \eqref{eq:rg_u}. Finally the system is discretized with a sampling time of $T_s=0.05$ s, yielding
\begin{equation}\label{eq:dt_system}
    x(t+1)=\begin{bmatrix}
    0.9701 & 0.0459\\
   -1.1610  &  0.8312
    \end{bmatrix}x(t) + \begin{bmatrix}
    0.0299 \\
    1.1610
    \end{bmatrix}v(t).
\end{equation}

\begin{figure}[t]
    \centering
    \begin{overpic}[width=.8\linewidth]{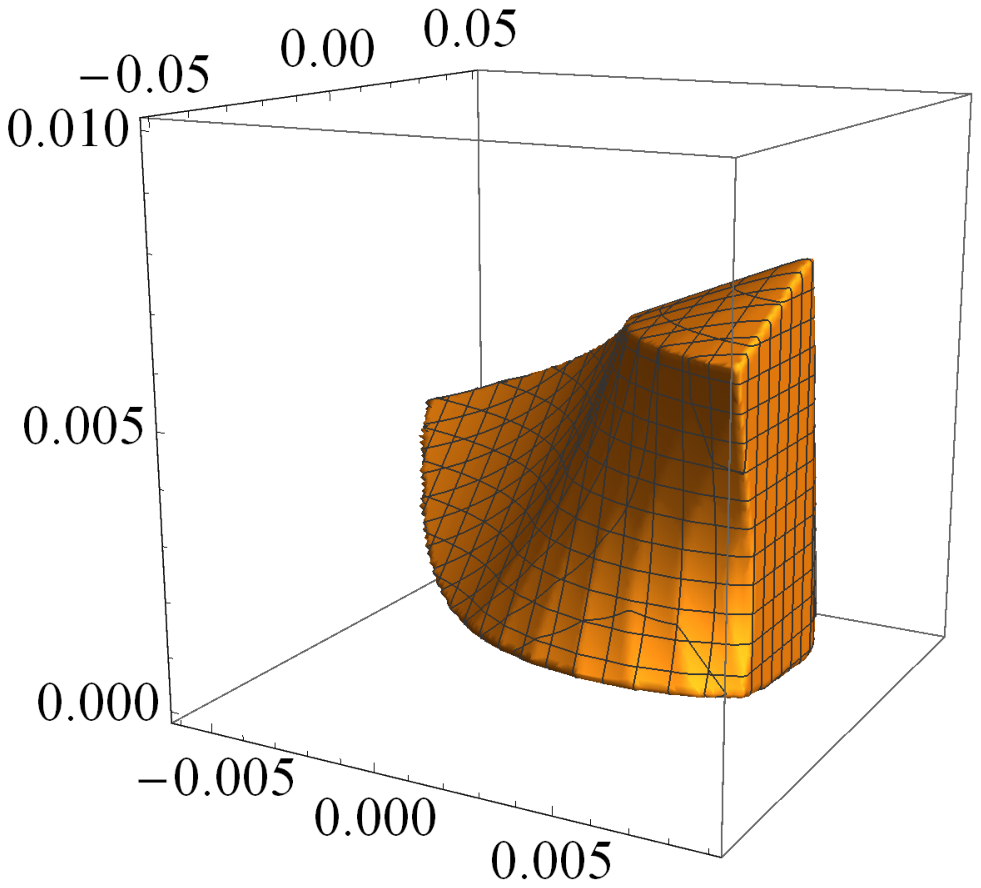}
    \put (13,50){$v$}
    \put (27,0){$x_1$}
    \put (25,80){$x_2$}
    \end{overpic}\\
    \vspace{1cm}
    \begin{overpic}[width=.8\linewidth]{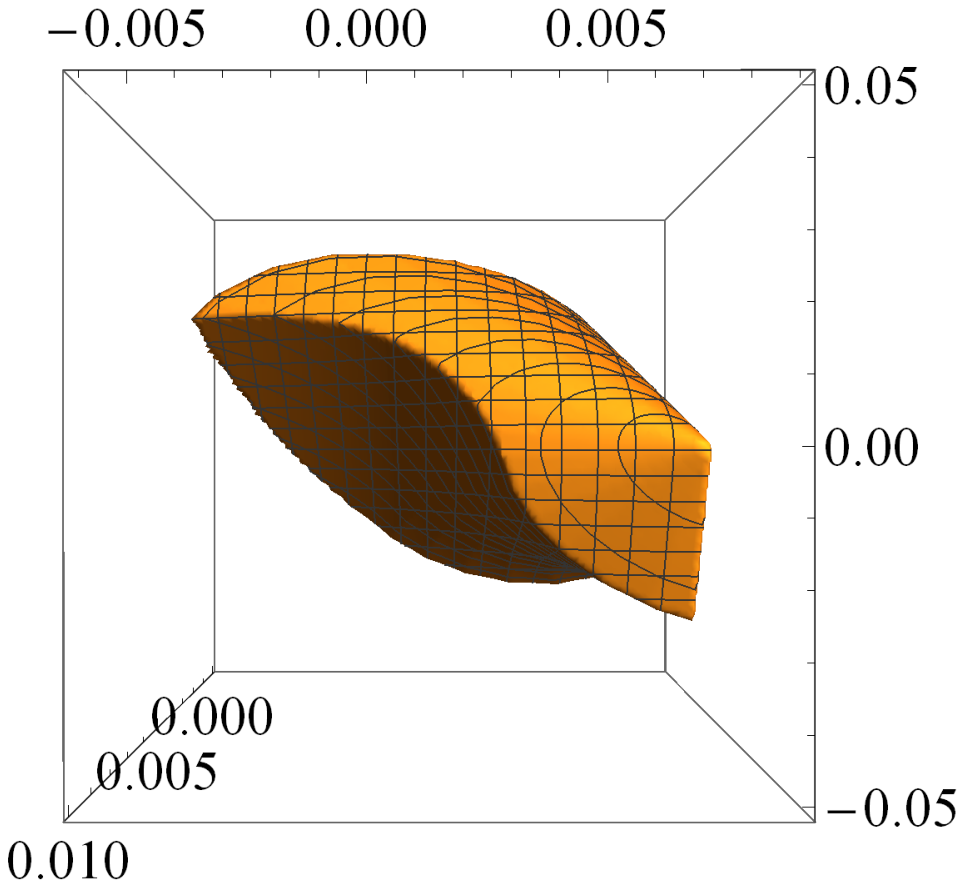}
    \put (47,80){$x_1$}
    \put (82,50){$x_2$}
    \put (17,20){$v$}
    \end{overpic}
    \vspace{1cm}
    \caption{$\tilde{\mathcal{O}}_\infty$ for system \eqref{eq:dt_system} viewed from two different angles.}
    \label{fig:oinf}
\end{figure}

The resulting $\tilde{\mathcal{O}}_\infty$ is depicted in \figurename~\ref{fig:oinf}. For its computation we considered $\varepsilon=10^{-2}$.
$\tilde{\mathcal{O}}_\infty$
was finitely determined after $k^\ast=35$ iterations and it is described by 87 inequalities. The elapsed time to compute $\tilde{\mathcal{O}}_\infty$ was 265 s. Out of the total 105 inequalities resulting from this horizon, 18 were determined to be redundant. It must be remarked that the upper bound $\bar{k}$ computed using the Lyapunov approach as in \cite{bemporad1998reference} would give an unreasonably long horizon of $\bar{k}=786$ and that, in absence of a sound methodology, in previous publications reporting  this example the horizon $k$ was estimated empirically. 

\subsubsection*{Application 2: Satellite Rendezvous}

In this example we apply the proposed methodology to compute $\tilde{\mathcal{O}}_\infty$ for a deputy-chief satellite rendezvous.
In this setting, the control objective is for the deputy satellite to approach the chief while staying in line of sight \cite{kalabic2011reference}. A non-inertial Hill frame is attached to the chief spacecraft where the three axes are the radial direction towards earth, the along-track direction towards the chief spacecraft, and the cross-track direction along the chief’s angular momentum vector, respectively.

The orbit can be represented by the following linearized model,
\begin{multline}\label{eq:sys_satellite}
\dot{x}(t)=\begin{bmatrix}
0 & 0 & 0 & 1 & 0 & 0\\
0 & 0 & 0 & 0 & 1 & 0\\
0 & 0 & 0 & 0 & 0 & 1\\
3\nu^2 & 0 & 0 & 0 & 2\nu & 0\\
0 & 0 & 0 & -2\nu & 0 & 0\\
0 & 0 & -\nu^2 & 0 & 0 & 0
\end{bmatrix}x(t)\\
+\begin{bmatrix}
0 & 0 & 0\\
0 & 0 & 0\\
0 & 0 & 0\\
1 & 0 & 0\\
0 & 1 & 0\\
0 & 0 & 1\\
\end{bmatrix}u(t),
\end{multline}
where $\nu=0.0011$ rad s$^{-1}$, $x=[x_1~x_2~x_3~\dot{x}_1~\dot{x}_2~\dot{x}_3]^T$ are the relative positions and velocities along the three axes of the Hill frame, and $u=[F_1~F_2~F_3]^T$ are the three components of the thrust in the aforementioned axes. System \eqref{eq:sys_satellite} is subject to the following constraints:
\begin{equation}
\begin{aligned}
(\tan ^2 \gamma) (x_2+0.01)^2 - x_1^2 - x_3^2 \geq 0 \\
u^2_{max} - u_1^2 -u_2^2 - u_3^2 \geq 0\\
    x_2 \geq 0
\end{aligned}
\end{equation}
with $\gamma=15^\circ$ and $u_{max}=4~\textnormal{N}$.
Discretizing \eqref{eq:sys_satellite} with $T_s=0.5~\textnormal{s}$ and controlling it with the control law $u(t)=Fx(t) + Gv(t)$, with LQR gain $F$ and feedforward gain $G$ such that $v$ becomes the reference position of the deputy spacecraft yields
\arraycolsep=0.25pt
\begin{equation}\label{eq:sys_satellite_final}
    x(t+1)=Ax+Bv,
\end{equation}
where
\begin{equation*}
A\!=\!\!\begin{bmatrix}
0.5698 & 0 & 0 & 0.1721 & 0 & 0\\
0 & 0.9121 & 0 & -0.002 & 0.3517 & 0\\
0 & 0 & 0.5698 & 0 & 0 & 0.1721\\
0.0004 & -0.3517 & 0 & -0.0005 & 0.4069 & 0\\
    0 & 0 & -1.7207 & 0 & 0 & -0.3117
\end{bmatrix}
\end{equation*}

\begin{figure}
    \centering
    \begin{overpic}[width=.7\linewidth]{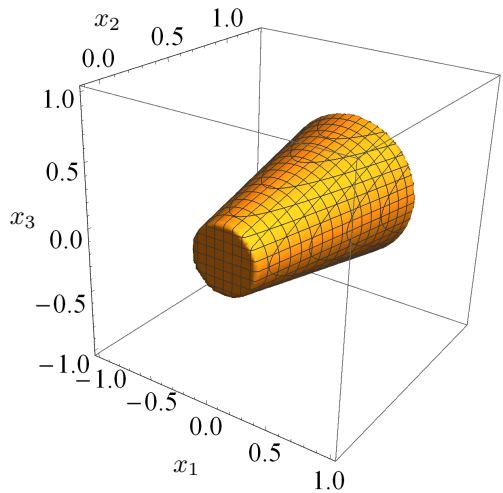}
    \put(28,5){$x_1$}
    \put(12,95){$x_2$}
    \put(-5,55){$x_3$}
    \end{overpic}
    \vspace{.3cm}
    \caption{3D slice of the 9-dimensional $\Oinf$ of system \eqref{eq:sys_satellite_final} corresponding to $x_4=x_5=x_6=v_1=v_3=0$, $v_2=0.05$.}
    \label{fig:oinf_satellite}
\end{figure}

\arraycolsep=5pt
\begin{equation*}
B=\begin{bmatrix}
    0.0363  &  0.0001 &   0\\
   -0.0001 &   0.1777  &  0\\
   0  & 0 &   0.0363\\
    0.1453   & 0.0005  &  0\\
   -0.0003  &  0.7108   & 0\\
    0  & 0    &0.1453\\
\end{bmatrix}.
\end{equation*}
A 3D slice of the $\Oinf$ computed applying Algorithm \ref{alg:algorithm2} is depicted in \figurename~\ref{fig:oinf_satellite}. For this system $k^\ast$ was determined to be $k^\ast=13$ and $\Oinf$ is defined by 31 inequalities.  This represents a very significative decrease compared to the 30 step prediction horizon (and the resulting 90 constraints) which were empirically estimated in \cite{kalabic2011reference}. The computation of $\tilde{\mathcal{O}}_\infty$ took 414 s.

\section{Concluding remarks}
In this paper we presented an approach for the open problem of eliminating redundant polynomial constraints in a semi-algebraic set. The approach is based on Sum of Squares arguments. The paper also presented some important applications of the method to constrained control. In particular we showed that the method can be used in an effective way to reduce the number of constraints in online implementations of Model Predictive Control and Reference Governor schemes. Interestingly, the method can also be used to systematically eliminate the terminal constraint in MPC formulations.
\bibliographystyle{IEEEtran}
\bibliography{bib.bib}{}
\appendix

\section{Krivine--Stengle Positivstellensatz}\label{sec:psatz}
Before presenting the Krivine--Stengle Positivstellensatz, a few definitions need to be introduced. For the sake of simplicity, the following concepts will not be explained in depth, and will rather be mathematically characterized. For further information on the matter, the reader is referred to \cite{parrilo2000structured,bochnak2013real,stengle1974nullstellensatz} and the references therein.

\begin{definition}{(Multiplicative Monoid)}
The multiplicative monoid generated by a set of polynomials $P=\{p_i\}_{i=1}^m$, $p_i\in\mathbb{R}[x_1,\ldots,x_n],~ i=1,\ldots,m$ is the set of finite products of the $p_i$, including the unity:
\begin{equation*}
    \textnormal{Monoid}(P)=\left\{\prod_{i=1}^m p_i^{k_i},\  k_i\in\mathbb{Z}_{\geq 0},~ i=1,\ldots,m\right\}.
\end{equation*}
\end{definition}
\begin{definition}{(Cone)}
The cone generated by a set of polynomials $P=\{p_i\}_{i=1}^m$ , $p_i\in\mathbb{R}[x_1,\ldots,x_n],~i=1,\ldots,m$ is the sum of the elements of $\textnormal{Monoid}(P)$ multiplied by some sum of squares polynomials $s_i$:
\begin{multline*}
    \textnormal{Cone}(P)=\left\{ s_0 + \sum_i s_i g_i : s_i \in \Sigma[x_1,\ldots,x_n], \right. \\ g_i \in \textnormal{Monoid}(P) \Bigg\}.
\end{multline*}
\end{definition}
\begin{definition}{(Ideal)}
The ideal generated by a set of polynomials $P=\{p_i\}_{i=1}^m$, $p_i\in\mathbb{R}[x_1,\ldots,x_n],~i=1,\ldots,m$ is the sum of the products of the $p_i$ and some polynomials $t_i$:
\begin{equation*}
    \textnormal{Ideal}(P)=\left\{\sum_{i=1}^m t_i p_i:  t_i\in\mathbb{R}[x_1,\ldots,x_n],~i=1,\ldots,m \right \}.
\end{equation*}
\end{definition}
At this point, the Krivine--Stengle Positivstellensatz can be expressed as follows:
\begin{theorem}{(Krivine--Stengle Positivstellensatz)}
Let $f_i(x)$, $i\in\mathcal{I}$, $g_j(x),\ j\in\mathcal{J}$, $h_k(x),\ k\in\mathcal{K}$ be finite sets of polynomials in $\mathbb{R}[x]$, $C=\textnormal{Cone}(f_i)$, $M=\textnormal{Monoid}(g_j)$, and $I=\textnormal{Ideal}(h_k)$, then the set
\begin{equation*}
    \{x : f_i(x)\geq 0,i \in \mathcal{I},g_j(x)\neq\ 0,j \in \mathcal{J}, h_k(x)=0, k \in \mathcal{K}\}
\end{equation*}
is empty if and only if
\begin{equation*}
    \exists f\in C,\,g\in M,\,h\in I\ :\ f+g^2+h=0.
\end{equation*}
\end{theorem}

\end{document}